\documentclass[aps,pra,amsmath,preprint,amssymb,groupedaddress,superscriptaddress]{revtex4-2}
\usepackage{graphicx}
\usepackage{hyperref}
\usepackage{natbib}
\usepackage[utf8]{inputenc}
\graphicspath{{Figures/}}
\begin{document}
	\title{Interactions of Ions and Ultracold Neutral Atom Ensembles in Composite Optical Dipole Traps: Developmentsand Perspectives}
	\author{L. Karpa}
	\email[]{karpa@iqo.uni-hannover.de}
	\affiliation{Leibniz Universität Hannover, Institut für Quantenoptik, 30167 Hannover, Germany}
	\date{\today}
\begin{abstract}
Ion--atom interactions are a comparatively recent field of research that has drawn considerable attention due to its applications in areas including quantum chemistry and quantum simulations. In first experiments, atomic ions and neutral atoms have been successfully overlapped by devising hybrid apparatuses combining established trapping methods, Paul traps for ions and optical or magneto-optical traps for neutral atoms, respectively. Since then, the field has seen considerable progress, but the inherent presence of radiofrequency (rf) fields in such hybrid traps was found to have a limiting impact on the achievable collision energies. Recently, it was shown that suitable combinations of optical dipole traps (ODTs) can be used for trapping both atoms and atomic ions alike, allowing to carry out experiments in absence of any rf fields. Here, we show that the expected cooling in such bichromatic traps is highly sensitive to relative position fluctuations between the two optical trapping beams, suggesting that this is the dominant mechanism limiting the currently observed cooling performance. We discuss strategies for mitigating these effects by using optimized setups featuring adapted ODT configurations. This includes proposed schemes that may mitigate three-body losses expected at very low temperatures, allowing to access the quantum dominated regime of interaction.
\end{abstract}
\maketitle
\section{Introduction}
The study of interactions between neutral atoms and ions is a highly topical and promising area of research at the intersection of several disciplines including atomic physics, chemistry and quantum simulations~\cite{Smith2005,Grier2009,Schmid2010,Zipkes2010,Ravi2012,Hall2012,Rellergert2011,Haerter2014,Haze2018,Meir2016,Tomza2019,Deiglmayr2012}. Many current experiments exploit combinations of established trapping methods, {building on the concept of ion--atom hybrid traps first demonstrated by Smith et al. in 2005}~\cite{Smith2005}, and have revealed interesting effects and phenomena such as sympathetic cooling~\cite{Zipkes2010,Ravi2012,Dutta2017,Haze2018,Sivarajah2012,Haerter2014,Tomza2019,Feldker2020,Rellergert2013,Goodman2012,Weckesser2021b}, state-to-state chemistry~\cite{Wolf2017}, {non-Maxwellian }and superstatistical energy distributions~\cite{Rouse2017,Chen2014}, charge {exchange and} transfer~\cite{Dieterle2021,Smith2014}, {and ion--atom Feshbach resonances}~\cite{Weckesser2021b}. Despite this rapid progress, one of the problems encountered in most experiments suitable for studying generic combinations of ion and atom species is related to the fact that they employ radiofrequency (rf) fields to confine ions. The related limitations to the achievable collision energies can be traced back to the presence of rf fields. It was shown that the latter inherently invoke so-called micromotion-induced heating ~\cite{Cetina2012}, a mechanism limiting the range of accessible collision energies to millikelvins and higher~\cite{Tomza2019}, even if the ions are prepared in the motional ground state~\cite{Meir2016}. 

At the same time, reaching the regime of interaction where quantum effects dominate likely requires cooling to a threshold several orders of magnitude lower, determined by the individual combination of ions and atoms~\cite{Tomza2019,Cetina2012}. In the rapidly developing field, several promising approaches have been brought forward or demonstrated. For example, controlled Rydberg excitation can be used to achieve very low kinetic energies of ions positioned in an atomic cloud~\cite{Kleinbach2018,Schmid2018,Dieterle2021}. Another approach that was successfully applied to achieve sympathetic cooling close to the quantum regime makes use of the fact that the impact of heating depends on the atom-ion mass ratio and can be mitigated by immersing ions with a large mass, e.g., Yb$^+ $ in a cloud of extremely light neutral atoms such as Li~\cite{Feldker2020}. Despite these achievements, a generic scheme for combining ions and atoms in a way that allows for observing ultracold interactions in the quantum regime has been a sought-after goal in the field. 

Here, we discuss a recently demonstrated realization of a trap allowing to combine atomic ions and neutral atoms in absence of rf fields, based on confinement provided by optical potentials~\cite{Schneider2010,Lambrecht2017,Schmidt2018,Schaetz2017,Karpa2019,Schneider2012,Cormick2011}. In principle, this approach is applicable to any combination of ions and atoms with a finite polarizability, as is the case for all commonly used ionic and atomic species including, for example, Yb$ ^{+} $, Ba$ ^{^+} $, Ca$ ^{+} $, Mg$ ^{+} $, Be$ ^{+} $ ions and Li, K, Na, Rb, Yb, Er atoms~\cite{Karpa2019}. For Ba$ ^{+} $ overlapped with Rb, a system that has been studied in great detail using conventional rf-based ion traps~\cite{Haerter2012,Haerter2013a,Kruekow2016}, rf-free trapping was shown to enable efficient sympathetic cooling in the ultracold regime well below the Doppler limit by avoiding micromotion-induced heating. We first summarize the main methods and findings of the reported work. We then numerically investigate the impact of beam overlap on the predicted cooling performance. Lastly, we discuss methods for mitigating such detrimental effects as well as schemes that may allow to observe elastic ion--atom collisions in a temperature range where losses from three-body recombination are expected to dominate~\cite{Kruekow2016a}. 
\section{Materials and Methods}
%
\subsection{Ultracold Ion--Atom Interactions in Optical Dipole Traps}

\begin{figure}[h!]
		\includegraphics[width=0.45\textwidth]{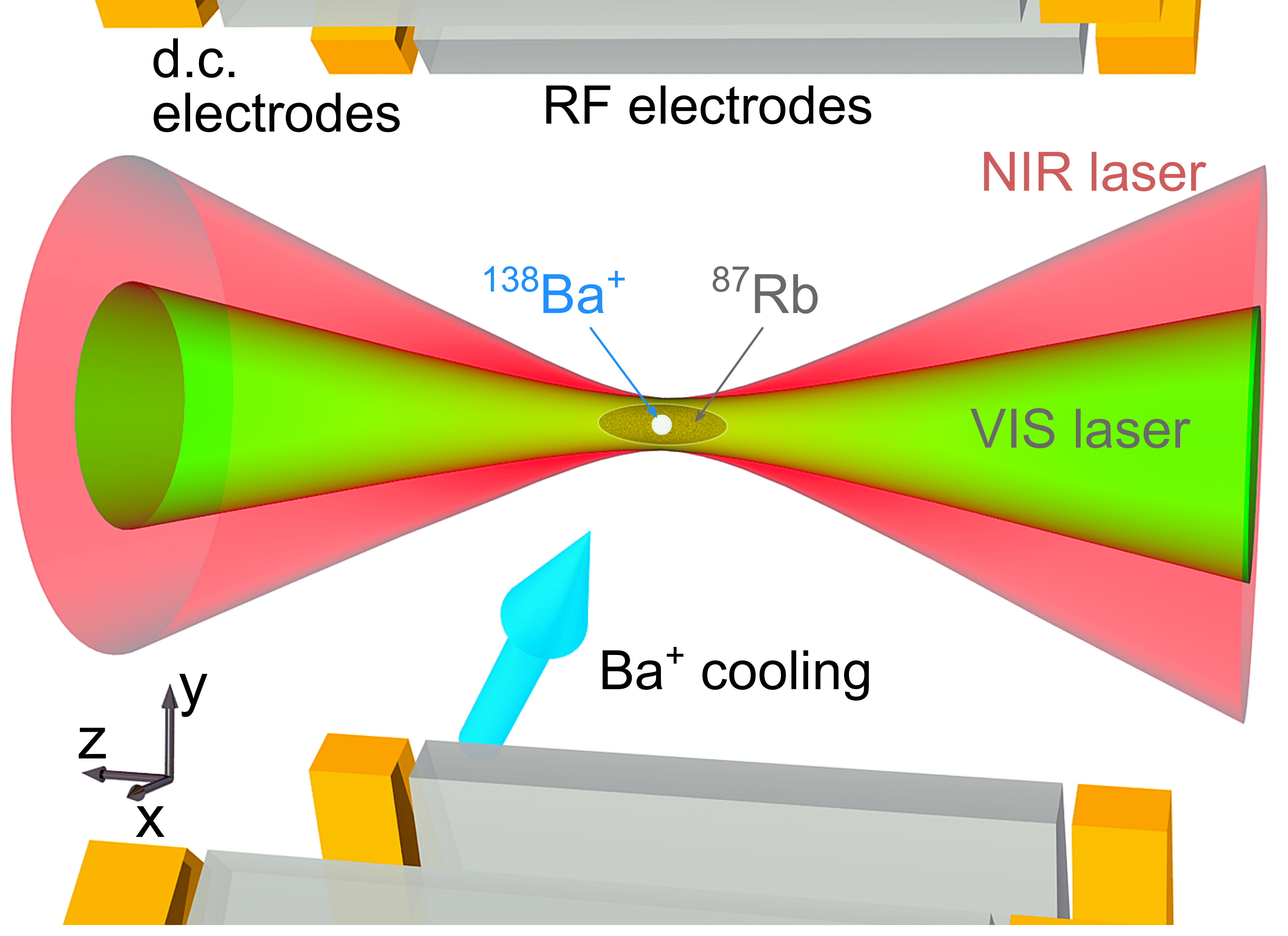}
	\caption{Schematic of the experimental setup. Bichromatic dipole traps (biODT) comprised of two lasers, VIS and NIR, are used to simultaneously trap a $ ^{\text{138}}\text{Ba}^+ $ ion and a cloud of ultracold $^{\text{87}}\text{Rb}$ atoms. From Schmidt, Weckesser, Thielemann,	Schaetz, and Karpa (2020) ~\cite{Schmidt2020}.
	}	
	\label{fig:setup4cooling}
\end{figure}

\begin{figure}[b]
		\includegraphics[width = 0.65 \textwidth]{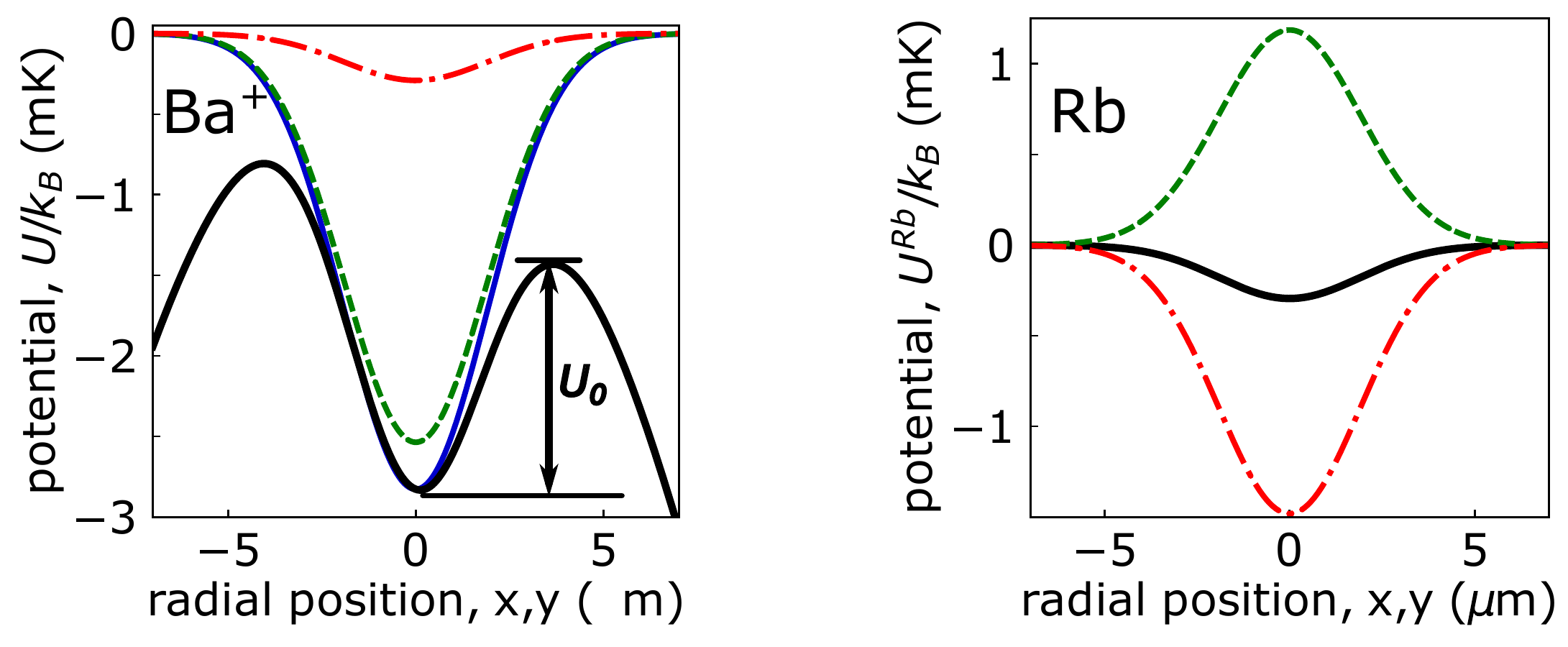}
	\caption{
		(\textbf{left}) Calculated optical potential for $ \text{Ba}^+ $ at $ z = \text{0} $ (blue, solid) with VIS (green, dashed) and NIR (red, dash-dotted) contributions. Taking into account electrostatic defocusing and stray fields (black, thick) yields an effective trap depth of $ U_0 / k_B \approx \text{1.4} ~ \text{mK} $.  (\textbf{right}) Calculated bichromatic potential for $ \text{Rb}$ (solid, black) with a trap depth $ U_0^{\text{Rb}} / k_B \approx 300 ~ \mu \text{K} $. From Schmidt, Weckesser, Thielemann,	Schaetz, and Karpa (2020) ~\cite{Schmidt2020}.
	}
	\label{fig:bichro}
\end{figure}

Simultaneous optical trapping of atoms and ions is a comparatively recent approach for avoiding radiofrequency fields~\cite{Schneider2010}. 	
Optical traps provide an extremely versatile and powerful tool-set that has lead to several breakthroughs in trapping and manipulating neutral atom ensembles over the coarse of several decades~\cite{Grimm2000,Bloch2008}. In the case of ions however, Coulomb interactions with residual stray electric fields or with other ions typically dominate over the much smaller optical dipole forces~\cite{Schaetz2017,Karpa2019}. Nonetheless, it was shown that with adapted methods for preparing ions, such as optimized geometries and improved compensation of stray electric fields, essentially the same methods can be applied to realize optical trapping of single ions~\cite{Schneider2010}, Coulomb crystals~\cite{Schmidt2018}, long lifetimes~\cite{Lambrecht2017} and state-selective potentials~\cite{Weckesser2021}. In general, the forces exerted by a single optical field on the ions and the atoms are different and can even have opposite signs. For example, this is the case for barium ions and rubidium when exposed to a laser beam operated at a wavelength of $\lambda_{VIS} = 532 $ nm, as illustrated in Figure \ref{fig:setup4cooling}. However, the potentials can be tuned by introducing an additional optical field with a wavelength of \mbox{$\lambda_{NIR} =$ 1064 nm} which allows to create overall attractive bichromatic potentials as illustrated in Figure \ref{fig:bichro}. Since the optical potentials from both ODT lasers are attractive for Ba$ ^{+} $, the ion is largely insensitive to fluctuations of the relative ODT beam positions on the order of a beam waist radius $ w_0 $. In contrast, Rb is repelled from  regions of high intensity of the VIS beam which has to be overcome by the NIR ODT at $\lambda_{NIR}$. Therefore, the bichtomatic trap (biODT) for Rb, and consequently the overlap between ions and atoms, can be drastically altered by fluctuations of the same magnitude, as we will show quantitatively later on. 

The experimental setup shown in Figure \ref{fig:setup4cooling} comprises a set of high-power lasers and a conventional linear Paul (rf) trap used for preparing a single ion, as well as for its detection after completion of the interaction phase during which the Paul trap is turned off. The preparation of the atomic sample follows standard techniques such as magneto-optical trapping, transferring the atoms to an optical dipole trap and evaporative cooling therein. Optical trapping of ions is achieved by employing previously developed techniques~\mbox{\cite{Schneider2010,Schaetz2017,Huber2014,Lambrecht2017,Schmidt2018,Karpa2019}}. Trapping probabilities $ p_{opt} \approx 1$ on timescales of ms relevant to the work discussed here are routinely achieved even if additional techniques for extending the lifetime, e.g., repumping from electronic states that experience a repulsive optical force~\cite{Lambrecht2017}, are omitted in favour of reducing the number of optical fields that might interact with the neutral atom ensemble. \\

\begin{figure}[h!]
		\includegraphics[width = 0.55\textwidth]{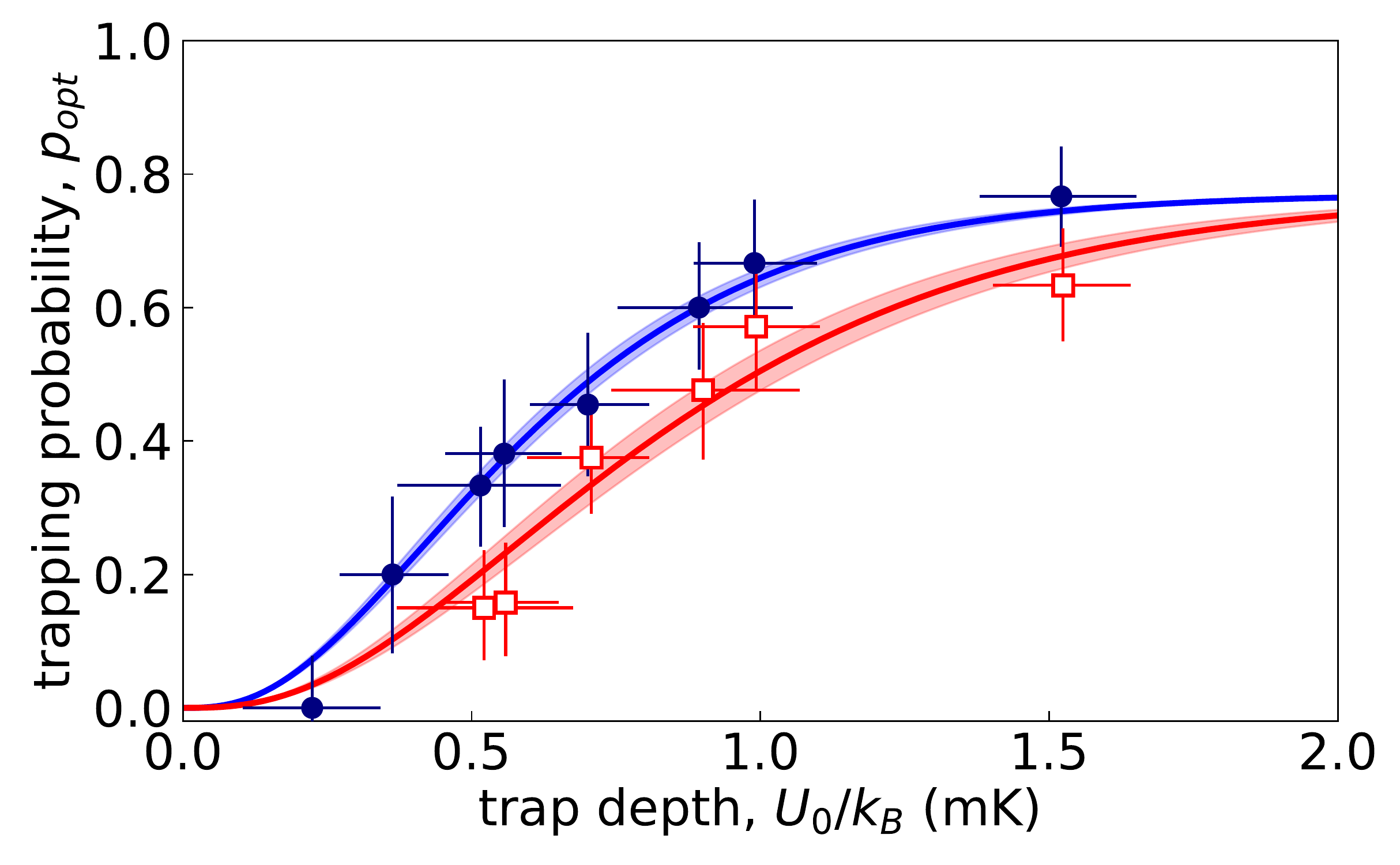}
	\caption{
		Sympathetic cooling of a $ \text{Ba}^+ $ ion in a cloud of ultracold $ \text{Rb}$ atoms. Open squares: experimental data taken after placing the ion in the bichromatic trap in absence of atoms. A fit with a modified radial-cutoff model\,\cite{Schneider2012} to the data (lower solid line) yields a temperature of $ T_{init}^{\text{Ba}^+} = \text{357} \pm \text{22} \, \mu \text{K} $. Full circles: the same experiment carried out with atoms, yielding $ T_{symp}^{\text{Ba}^+} = \text{259} \pm \text{10} \, \mu \text{K} $ (upper solid line). From Schmidt, Weckesser, Thielemann,	Schaetz, and Karpa (2020) ~\cite{Schmidt2020}. 
	}
	\label{fig:cooling}
\end{figure}

Following this approach, it was shown that overlapping a Ba$ ^{+} $ ion initially prepared in the Paul trap and Doppler cooled to approximately $ T_D = 365 \, \mu \text{K}$ with a small ensemble of about 500 ultracold $ ^{87} \text{Rb} $ atoms with a temperature of roughly $ 30 \, \mu \text{K} $ in a bichromatic trap, leads to a significantly improved optical trapping probability as shown in Figure \ref{fig:cooling}.  According to the radial-cutoff model~\cite{Tuchendler2008,Schneider2012} which allows to extract the temperature of single atoms by repeatedly measuring the trapping probability for different trap depths $ U_0 $, such an increase of $ p_{opt} $ is an unambiguous signature of cooling. The corresponding measured change of mean kinetic energy amounted to $ \Delta E_{kin}  = - k_B ~ (98 \pm 24) ~ \mu \text{K} $, where $ k_B $ denotes the Boltzmann constant. In a reference measurement where the atoms were removed in the bichromatic trap, the ion's temperature remained close to its initial value of $ T_D$, showing that the interaction with the ultracold cloud enabled sympathetic cooling to temperatures that are inaccessible in experiments where the ions are confined with rf fields. While in absolute terms, the observed cooling effect was moderate, the currently achievable performance could be limited by systematic effects, in particular fluctuations of the overlap between the two dipole trap beams. We will discuss the expected impact of such effects in the following.

\section{Results}
\subsection{Impact of fluctuating Dipole Trap Alignment}
While recent experiments in bichromatic traps have demonstrated the onset of sympathetic cooling, as summarized in Figure \ref{fig:cooling}, one open question is if the employed method is suitable for achieving thermal equilibration of the ion with the surrounding neutral atom gas, ultimately allowing for entering the s-wave scattering regime. In addition to the absolute measurement of the temperature performed in~\cite{Schmidt2020}, it is therefore instructive to analyse the expected level of performance when taking into account several technical limitations specific to the employed experimental setup. The basis for such an analysis are the following observations:
\begin{enumerate}
	\item All results reported in~\cite{Schmidt2020} were obtained by averaging over several experimental realizations.
	
	\item Individual realizations showed substantial fluctuations of ODT overlap. That is, the atoms were aligned with the center of the Paul trap in about every third realization which was observed using absorption imaging in reference measurements carried out without Ba$ ^{+} $ ions. Since the imaging of the atomic cloud was not part of the experimental sequence, it was not possible to post-selectively restrict the analysis to the cases where overlap was detected. Consequently, the measurements reported in~\cite{Schmidt2020} represent averaging over situations where the ions are overlapped with the atoms and those where the ion--atom overlap was strongly reduced or negligible.
	
	\item An increase of the ion--atom interaction time in the biODT did not lead to a significant enhancement of optical trapping probabilities. Seemingly, this would imply that the cooling is not improved beyond the level shown in Figure \ref{fig:cooling}.
	
	\item Repeating the experiment with different experimental parameters such as average trap overlap and initial ion temperature resulting from long-term drifts robustly yielded the same change of the apparent temperature $ k_B \Delta E_{kin}  $ determined after the ion--atom interaction phase in the biODT within the reported uncertainties.
	
	\item Longer ramp-up durations of the VIS/NIR ODTs, that is, build-up of the bichromatic potential, lead to formation of Rb$ ^+ $ and Rb$ _2^+ $ parasitic ions. For example, with a doubled ramp-up duration (2 ms) we observe the occurrence of such events with a probability of about 0.1 to 0.2. The production rate strongly depends on the intensities of the dipole trap beams, predominantly that of the VIS ODT. 
\end{enumerate}

\begin{figure}[h!]
	\includegraphics[width=1.0 \textwidth]{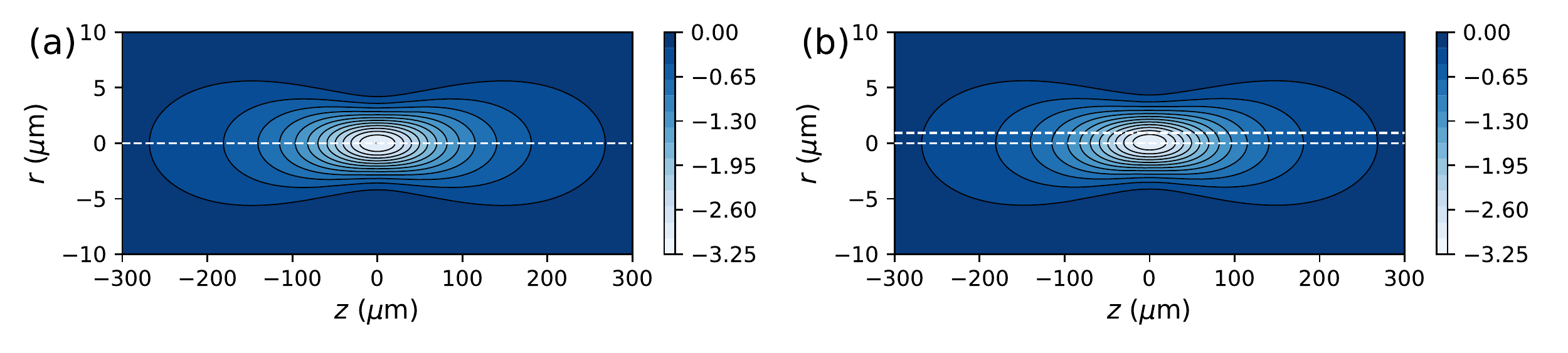}
	\includegraphics[width=1.0 \textwidth]{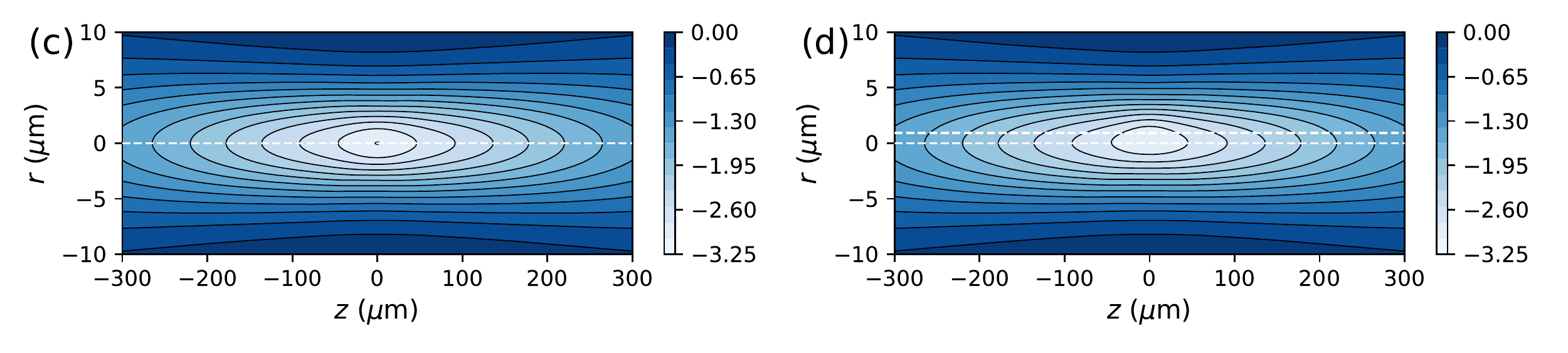}
	\includegraphics[width=1.0 \textwidth]{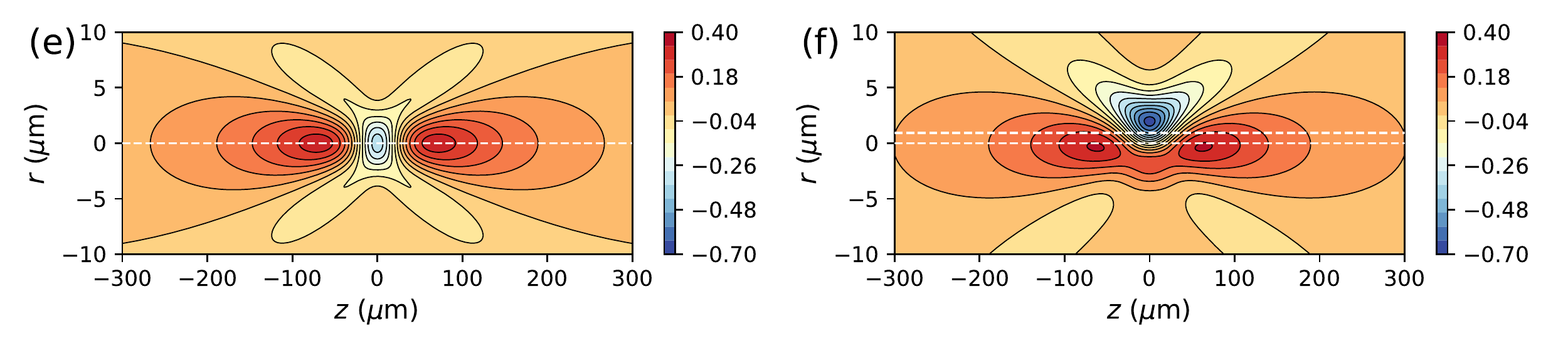}
	\includegraphics[width=1.0 \textwidth]{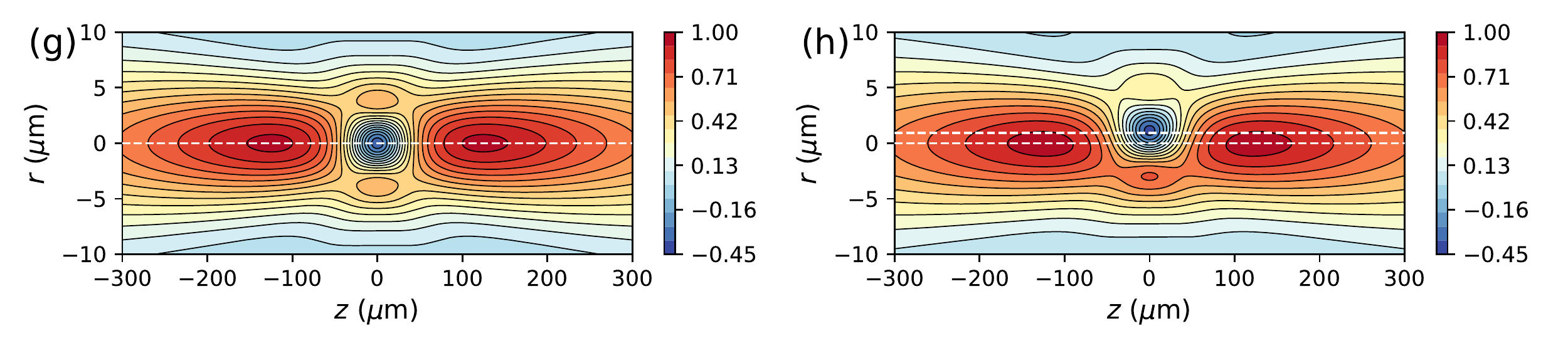}
\caption{{(\textbf{a}--\textbf{d}) }Calculated bichromatic potentials for $ ^{138} \text{Ba}^{+}$ and  {(\textbf{e}--\textbf{h})} for $ ^{87} \text{Rb}$. Potentials with (without) a relative radial displacement $ \Delta r = w_0/4 $ (indicated by the white dashed lines) between the NIR and VIS OTDs are depicted on the right (left).  {(\textbf{a},\textbf{b},\textbf{e},\textbf{f})} show results for matched $ 1/e^2 $ beam waist radii of $ w_0 = 3.7 \, \mu$m. The case where the VIS ODT waist is $ w_{0,VIS} = 2 w_0 $ is shown in {(\textbf{c},\textbf{d},\textbf{g},\textbf{h})}. The depth of the potentials $ U_{0}^{Ba,Rb}/k_B $ (in mK) is represented by false colours. For $ ^{138} \text{Ba}^{+}$, the effect of $ \Delta r $ on the potential is negligible. In contrast, for $ ^{87} \text{Rb}$ the same displacement drastically changes the potential landscape, shifting the trap minimum and increasing $ U_{0}^{Rb} $. The impact of the relative displacement is slightly mitigated for the configuration with $ w_{0,VIS} = 2 w_0 $.
		\label{calculatedODTdisplaced}}
\end{figure}

In order to estimate the impact of fluctuating relative positions of the individual optical traps on the rubidium ensembles, we calculate the bichromatic potentials assuming circular Gaussian profiles with a radial displacement $ \Delta r $ between the ODTs' optical axes as illustrated in Figure \ref{calculatedODTdisplaced}. Subsequently, we employ fitting around the harmonically approximated deformed bichromatic potentials to numerically obtain the radial locations of the new minima, potential depths, peak densities, as well as the effective densities as probed by ions that are still confined very close to the minimum of the VIS potential. It was assumed that the ODTs are perfect Gaussian beams without astigmatism or ellipticity and that the atomic ensembles were adiabatically adapting to the rearranged potentials, that is without additional heating or atom loss.  \\

\begin{figure}[h!]
	\includegraphics[width=1.0\linewidth]{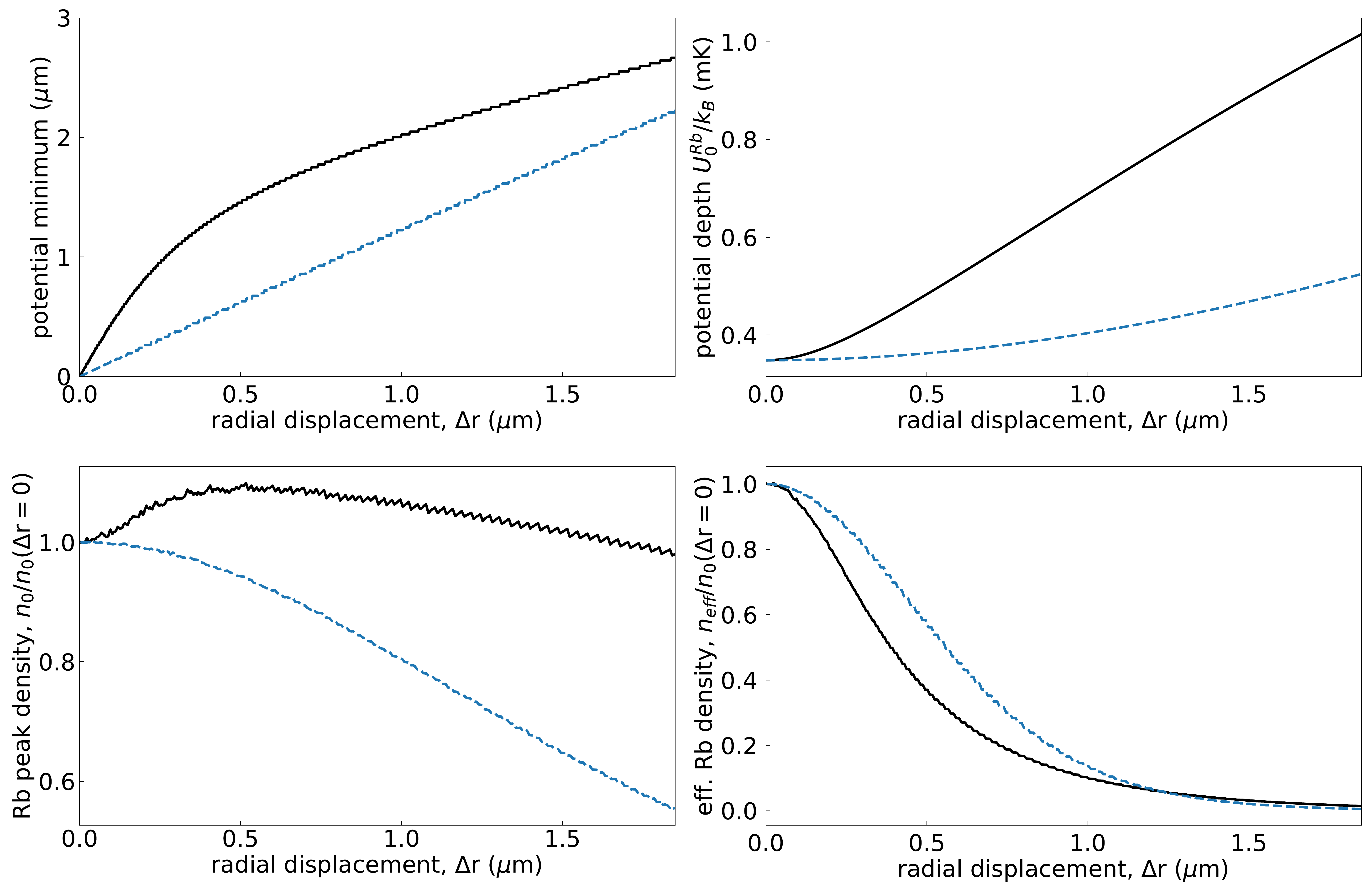}
	\caption{
		Numerically calculated locations of the minimum of the bichromatic potential for Rb atoms (\textbf{top left}), the corresponding trap depths $ U_{0}^{Rb} $ (\textbf{top right}), peak density normalized to the value obtained in the case of full VIS/NIR ODT alignment $ n_0/n_0(\Delta r = 0) $ (\textbf{bottom left}), and the effective normalized Rb density probed by the Ba ion $ n_{eff}/n_0(\Delta r = 0) $ as a function of radial displacement $ \Delta r $ between the VIS and NIR ODTs (\textbf{bottom right}). Results for OTDs with matched beam waist radii of $ w_0 = 3.7 \, \mu \text{m} $ ($ w_0^{VIS} = 2 w_0^{NIR} = 7.4 \, \mu \text{m} $) are shown as black solid lines (blue dashed lines). For the data in the lower panels a 10-point gliding average has been applied. \label{ODTdisplaced}}
\end{figure}

The corresponding results presented in Figure \ref{ODTdisplaced} show strong sensitivity with respect to radial parallel beam displacements $ \Delta r $, reflected in the drastic reduction of the effective Rb density at the position of the ion, even for displacements that are only a fraction of the ODT waist radius $ w_0^{VIS} = w_0^{NIR} = 3.7 \, \mu \text{m} $. {This is aggravated by the fact that the VIS and NIR components comprising the two bichromatic potentials affect the atoms and ions differently, as illustrated in Figure \ref{fig:bichro}. For Rb, both contributions are comparable in magnitude but opposite in sign, making the atoms a sensitive probe for beam displacement. In contrast, in the case of Ba$ ^{+} $, both potentials are attractive, with the contribution of the NIR field being small compared to that of the VIS component.} In the recently demonstrated bichromatic trap for ion--atom interactions~\cite{Schmidt2020}, the VIS and NIR ODT enter the vacuum chamber from opposite directions through high numerical aperture objectives used to focus the beams onto the ion. Therefore, differential fluctuations of the two opto-mechanical assemblies on the order of $ \mu \text{m}$, especially at frequencies in the acoustic range of $\sim 10^{3} $ Hz, are realistic. A straightforward strategy for reducing the impact of such fluctuations is to increase the radial extent of the VIS potential as compared to the NIR potential, effectively creating repulsive barriers around the atomic cloud. Indeed, our calculations show that a biODT configuration with a doubled VIS waist radius ($ w_0^{VIS} = 2 w_0^{NIR} = 7.4 \, \mu \text{m} $, dashed lines in Figure \ref{ODTdisplaced}), is less prone to changes resulting from radial ODT displacement. However, the remaining sensitivity still leads to a pronounced drop of the effective density for relative displacements of about $ 1 \, \mu \text{m} $, which is comparable to fluctuations of ODT pointing observed in previous measurements. Therefore, such shot-to-shot variations of the relative ODT alignment can explain the observation that the probability for preparing an atomic cloud at the position of the ion in an individual realization of the experiment is approximately $ p_{ovlp} \approx 0.3 \pm 0.1 $.\\ \vspace{-6pt}

\begin{figure}[h!]
		\includegraphics[width = 0.6 \textwidth]{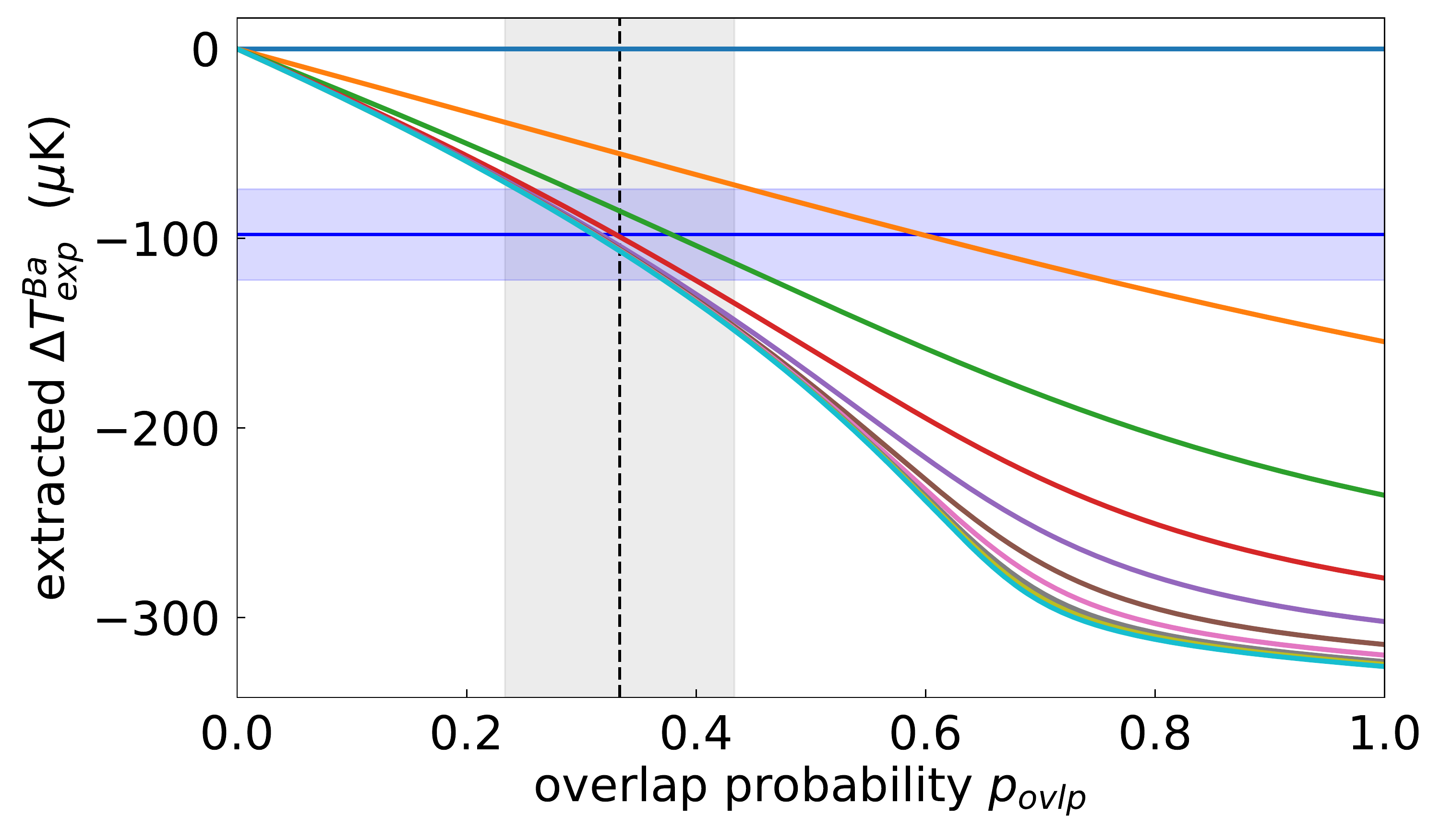}
	\caption{
		Expected changes of ion kinetic energies $ \Delta T^{Ba}_{exp} $ after $ N_{Lgvn}  = 0$ (top cruve) to $ N_{Lgvn} = 9$ (bottom curve) Langevin collision(s) for different assumed probabilities $ p_{ovlp} $ to observe full overlap between the ion and the atomic cloud stemming from fluctuations of the ODT alignment. The result from~\cite{Schmidt2020} is shown as the blue horizontal line with the shaded area denoting the uncertainties. The experimentally estimated overlap probability and its uncertainty are depicted as the black dashed line and the grey shaded region, respectively.
	}
	\label{fig:deltaTvsPalgn}
\end{figure}

An important question in view of these findings is how the observed intermittent ion--atom overlap affects the measured optical trapping probability of the ion $ p_{opt} $, as the observable used to derive its change of kinetic energy after a number of elastic collisions (between 0  and 9). To gain insight into the expected performance and limitations, we employ a simplified model assuming that the interaction between ions and atoms occurs with a probability of $ p_{ovlp} $, representing a situation where the full atomic peak density is probed by a point-like ion. We then calculate distributions of $ p_{opt} $ as a function of bichromatic trap depth for Ba$ ^{+} $ corresponding to an average over realizations where the ion temperature is $ T_D $ (weighted with $ 1 - p_{ovlp} $) and those where the ion has undergone $ N_{Lgvn} $ Langevin collisions (weighted with $ p_{ovlp} $). Subsequently, we perform fitting with a radial-cutoff model~\cite{Tuchendler2008,Schneider2012} to the obtained distributions in order to simulate the expected apparent temperature that a measurement and evaluation under these conditions would yield. The results of this simulation are shown in Figure \ref{fig:deltaTvsPalgn}.
The expected temperature in the case of full alignment (for $ p_{ovlp} = 1 $) shows sizeable changes of $ \Delta T_{exp}^{Ba} \sim - 100 \, \mu \text{K} $ that can be resolved within the experimental uncertainties of typical optical ion trapping experiments after the first and second collision (orange and green lines), decreasing as the ion approaches equilibrium with the surrounding atomic ensemble shown as the lower lying curves. The cooling reported in~\cite{Schmidt2020} is shown as the blue {horizontal} line. It is consistent with, \emph{ on average}, $ N_{Lgvn}  \approx 1 $, however assuming $ p_{ovlp} = 1 $. In contrast, taking into account the estimated significant overlap fluctuations, represented by the grey shaded region, 
reveals that the currently accessible observable $ p_{opt} $ yields very similar values of the apparent temperature for a wide range of $ N_{Lgvn} $ ranging from 1 to 9, with differences between these scenarios being comparable to the uncertainties of realistic temperature measurement experiments. This suggests that methods based on measuring the average $ p_{opt} $ may not be sensitive enough to distinguish between the onset of sympathetic cooling and near-complete equilibration in presence of fluctuations of ion--atom overlap at the currently achieved level. In combination, these considerations can explain the summarized observations, including the finding that a prolonged interaction duration apparently is not accompanied by enhanced sympathetic cooling: for realistic beam stabilities, the employed method to measure cooling by deriving $ p_{opt} $ from many experimental cycles is not sensitive enough to resolve the actual reduction in kinetic energy achieved in individual runs. Consequently, the cooling efficiency observed in~\cite{Schmidt2020} most likely establishes a lower bound on the actually achievable cooling that, in individual realizations of the experiment, may reach into the regime of thermal equilibration. In order to investigate the full performance of bichromatic traps in view of sympathetic cooling, such experiments require a setup that is more robust to differential position fluctuations. This could be realized, e.g, by overlapping the two ODT beams before they enter the vacuum chamber and by focusing them using the same objective. In this case, residual fluctuations would be common mode, such that they would no longer significantly affect the overlap between ions and atoms, although they still may result in additional heating.
\section{Discussion and Perspectives}
\subsection{Influence of Parasitic Ions on Sympathetic Cooling}
For the combination of Ba$ ^{+} $ and Rb, the presence of optical trapping fields can lead to the formation of parasitic ions, predominantly Rb$ ^{+} $ and Rb$ _2^{+} $~\cite{Haerter2013b}. These processes can release large amounts of energy compared to that of the unperturbed Ba$ ^{+} $, resulting in drastic heating. Some creation mechanisms can be attributed to resonantly enhanced multi-photon ionization (REMPI) in the atom ensemble, triggered by three-body recombination. They could be mitigated by using {lower atomic densities and} suitable ODT wavelengths to avoid molecular resonances~\cite{Haerter2013b}. However, this may not be possible in general, and, in the worst case, the parasitic ions have to be removed selectively without losing the target ion. During experimental phases where the ion is confined by rf fields, this can be achieved by employing a method based on parametric excitation~\cite{Landau1969} of parasitic ions, e.g., Rb$ ^{+} $ and Rb$ _2^{+} $, at twice their respective oscillation frequencies in the Paul trap~\cite{Vedel1990,Yu1993,Razvi1998,Haerter2013b}. Owing to the exponential increase of the kinetic energy characteristic for parametric modulation, this excitation scheme in conjunction with a high mass resolution of at least $ 1/138 $ allows for the selective removal of different barium isotopes from an ion crystal~\cite{Schmidt2020b} and is well suited for removing parasitic ions with a more distinct charge-to-mass ratio. With this technique, the trapping probability can be drastically improved, now allowing for efficient transfer into the bichromatic potential. \\

However, once the confinement of the ion is provided by optical potentials, the occurrence of parasitic ions will lead to loss, limiting the time available for ion--atom interactions. In the situation described in the previous section, complications due to increased probability for producing parasitic ions by photoionization arise when performing ion--atom collision experiments with durations longer than 0.5 ms. The expected improvement in sympathetic cooling would then be masked by the loss of ions in the cases where parasitic ions are produced, resulting in a decrease of $ p_{opt}$ and hence a higher apparent temperature. Apart from using sub-Doppler cooling of ions during the preparation~\cite{Monroe1995,Vuletic1998,Morigi2000,Karpa2013}, allowing to reduce the optical power required for trapping, this problem can be potentially avoided or mitigated by using an ODT operating in a wavelength range where photoionization is suppressed~\cite{Passagem2019,Haerter2013b}. In a more general context, the interaction of optical fields of high intensity with the atomic and molecular ensembles trapped therein is a highly topical and active area of research not only in view of ion--atom interactions, but also in the field of neutral gas quantum chemistry~\cite{Christianen2019}. The details of this interaction, and in particular its role in the formation of neutral molecular complexes, are still an open question and the subject of ongoing theoretical and experimental investigations~\cite{Gregory2020,Bause2021,Gersema2021}. The outlined modifications to ion--atom experiments may enable studies at sufficiently long interaction durations at least on the order of milliseconds, providing a complementary perspective on these interesting effects and phenomena. This approach is particularly promising in the case of fermionic neutral atoms, as in the $ ^{6}\text{Li}-{}^{171}\text{Yb}^+ $ system, where ultracold ion--atom interactions close to the quantum dominated regime are possible in the presence of intense optical fields for durations on the order of seconds~\cite{Feldker2020}.\\

Building upon advanced but well-established methods for controlling atoms, extending optical trapping to standing wave and lattice configurations~\cite{Morsch2006,Bloch2008,Grimm2000} opens an alternative route to avoiding this problem. Ultimately, experiments could be performed with a single ion or small ion crystals interacting with arrays of individually trapped atoms~\cite{Endres2016,Barredo2016,Barredo2018}. Since three-body collisions are inherently suppressed, ionization processes requiring high atomic densities such as REMPI would be inhibited. Additionally, blue-detuned optical potentials can be implemented in order to minimize exposure of the atoms to trapping light, as recently demonstrated in the case of neutral NaK molecules, where photo-assisted processes are a concern~\cite{Bause2021}. In the following, we propose and discuss two model configurations aiming to establish control over the overlap between ions and atoms such that the ratio of elastic collisions to three-body recombination events can be adjusted to allow for sympathetic cooling.
\subsection{Adapted Configurations for Experiments in the Ultracold Regime of Interactions}
\begin{figure}[h!]
	\includegraphics[width=0.55 \textwidth]{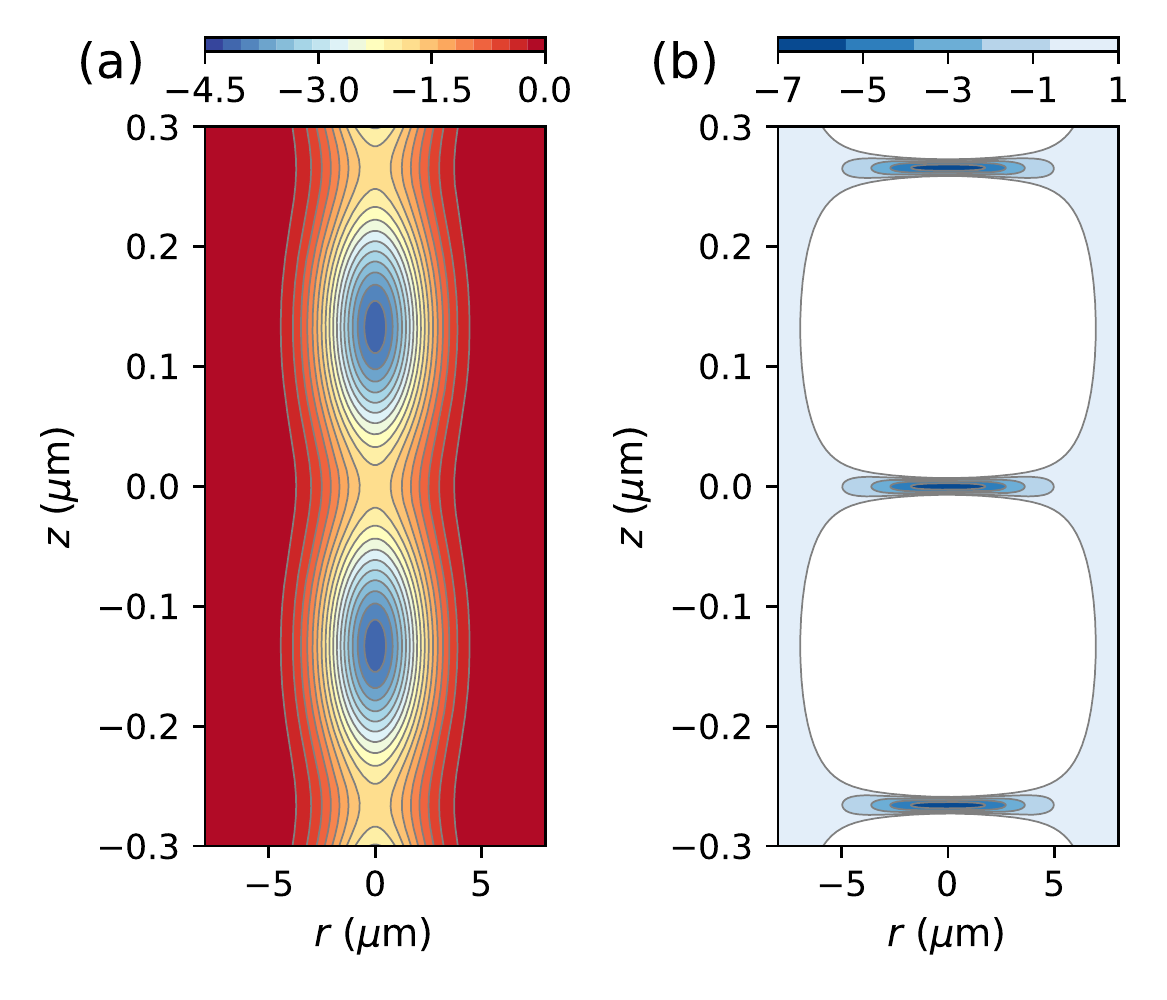}		
\caption{
		Calculated bichromatic optical potentials for $ ^{138} \text{Ba}^{+}$  {(\textbf{a})} and $ ^{87} \text{Rb}$ {(\textbf{b})} for a combination of a single NIR beam and a partially retro-reflected VIS OTD. The assumed reflected power is 5 $ \% $ of the input power $ P_{VIS}^{IN} = 469 \, \text{mW} $, the NIR ODT power is $ P_{NIR} = 106 \, \text{mW} $, all beams are focused to a $ 1/e^2 $ beam waist radius of $ w_0 = 3.7 \, \mu \text{m} $. The numeric values of the potentials $ U^{Ba}/k_B $ (in mK) and $ U^{Rb}/k_B $ (in $ \mu $K) are mapped to a false color representation. The potentials show the expected separation between the Ba$ ^{+} $ and Rb allowing for a drastic reduction of their wavefunctions' overlap. Since the effective Ba potential can be shifted with sub-wavelength precision by adjusting the electrostatic trapping potential while the potential for Rb remains unchanged, the ion--atom overlap can be dynamically varied, allowing to tune the elastic collision rate, as well as the ratio of elastic to inelastic~collisions.
		\label{LatticeBiODT}}
\end{figure}

In the future, with improved pointing and intensity stability of the trapping beams, entering the quantum regime of interaction for the chosen ion--atom combination is expected to be feasible but, in the current configuration, is likely to be highly challenging. This is largely owed to the fact that the associated energy limit corresponds to Rb temperatures below $ 100 ~ \text{nK} $~\cite{Cetina2012}. This usually involves the preparation of ultracold ensembles with typical atomic densities far in excess of $ 10^{14} ~ \text{cm}^{-3}$, such as a Bose--Einstein Condensate. Under such conditions, inelastic collision events, e.g., three-body recombination, are expected to occur several orders of magnitude more frequently than elastic ion--atom collisions that are responsible for sympathetic cooling~\cite{Kruekow2016,Krych2011}. Therefore, future experiments will have to provide a significantly improved control of the local atomic densities in the interaction region. This could be achieved in an extended experimental setup utilizing an axial standing wave of the VIS dipole trap, such that the ultracold rubidium atoms are repelled from the intensity maxima where the barium ion is held. An example for this configuration is shown in Figure \ref{LatticeBiODT}. A summary of the assumed experimental parameters is given in \mbox{Table \ref{LatticeTable}}.
\begin{table}[h!]
	\begin{tabular}{lcc}
		\toprule
		\textbf{}	& \textbf{$ ^{138} $Ba$ ^{+} $}	& \textbf{ $ ^{87} $Rb}\\
		\hline
		$P_{VIS / NIR} \, \text{(mW)}  $		& \multicolumn{2}{c}{ 469 / 106 } \\
		$w_0 \, ( \mu \text{m})  $		& \multicolumn{2}{c}{ 3.7 }  \\
		$\omega_{ax}/(2 \pi) \, \text{(kHz)}  $		& 860			& 1257\\
		$\omega_{rad}/(2 \pi) \, \text{(kHz)}  $		& 1.3			& 35\\
		$U_{0}/k_B \text{(K)} $ 		& $ 1.8 \times 10^{-3} $			& $ 6.5\times 10^{-6} $\\
		\toprule
	\end{tabular}
	\caption{
		Summary of properties of the calculated optical potentials for $ ^{138} \text{Ba}^{+}$ and $ ^{87} \text{Rb}$ as shown in Figure \ref{LatticeBiODT}: optical powers $P_{VIS / NIR} $, $ 1/e^{2} $ beam waist radii $w_0$, axial/radial trap frequencies $\omega_{ax/rad}$, and trap depths $U_{0}$. The power of the retroreflected VIS beam is $0.05 \,P_{VIS} $, sufficient to superimpose a standing-wave structure yielding high axial trapping frequencies.
		\label{LatticeTable}}
\end{table}

\begin{figure}[b]
\includegraphics[width=0.9 \textwidth]{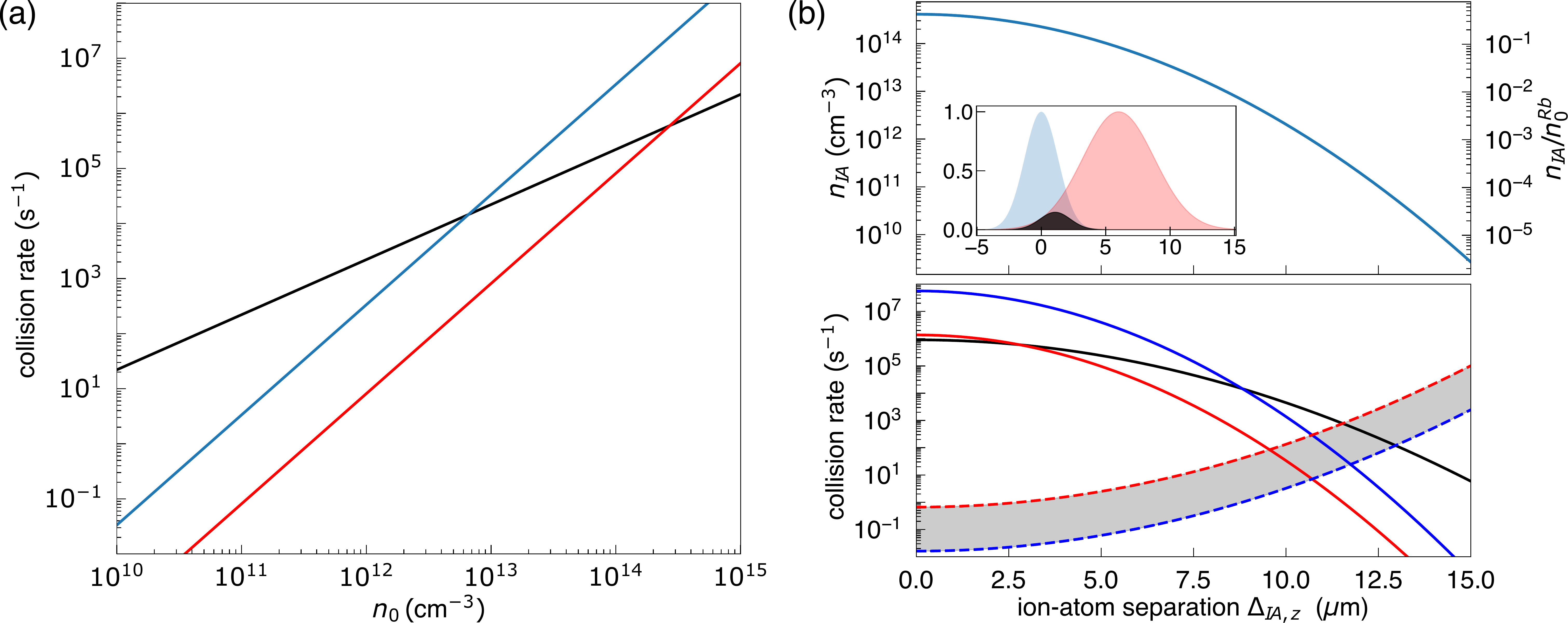}		
\caption{
 {(\textbf{a})} Calculated Langevin collision rate $ \gamma_{Lgvn} $ (black line) and the three-body recombination rate $ \gamma_{3b} $ for an ion with a kinetic energy of $ E_{kin} = k_B \times 370 \, \mu \text{K} $ (red line) and an ion with $ E_{kin} = k_B \times 1 \, \mu \text{K} $ (blue line).  {(\textbf{b})} Top panel: Effective (left axis) and normalized (to $ n^{Rb}_0 $) (right axis) rubidium density probed by a barium ion prepared at $ T_D \approx 370 \, \mu \text{K}$ (red shaded area) positioned at a distance of $  \Delta_{IA,z} $ from the center of an atomic cloud (blue shaded) cooled to $T_{at} = 1 \, \mu \text{K} $. The inset shows the calculated axial overlap (black area) for a separation of $ \Delta_{IA,z}  =  6 \, \mu \text{m}  $. Bottom panel: calculated Langevin (black solid line) and three-body recombination  (red and blue solid lines) rates, $ \gamma_{Lgvn}$ and $\gamma_{3b} $, respectively, for a Ba$ ^{+} $ ion prepared at $ T_D $ (red) and after assumed thermalization to $T_{at} $ (blue). The corresponding ratios $ \gamma_{Lgvn} / \gamma_{3b} $ are shown as red and blue dashed lines, with curves for intermediate ion temperatures lying in the grey shaded region.
	\label{Overlap}}
\end{figure}

This exemplary configuration gives rise to several important advantages afforded by the tight axial confinement of the ion. Firstly, its related high trapping frequencies allow for in situ Raman sideband cooling close to the motional ground state~\cite{Karpa2013}. For previously demonstrated bichromatic traps this is difficult because the attainable trap frequencies are not high enough to support cooling in the resolved sideband regime. Secondly, the overlap of atoms with regions of high intensity and with ions is reduced, manifesting in lower decoherence and three-body recombination rates and the reduced optical intensities required for trapping, whereas the lifetime of ions is increased~\cite{Lambrecht2017}. {For example, with an ensemble of $N_{Rb} \sim 10^{4} $ atoms, a Doppler cooled barium ion shifted into the center of the atomic cloud would experience a peak density of $ n_0 \sim 10^{15}  ~ \text{cm}^{-3}$, as shown in Figure \ref{Overlap}a, resulting in an expected loss through three-body recombination within less than $ 1  \, \mu \text{s}$, whereas the effective density for an ion left in the minimum of its bichromatic potential would be five orders of magnitude lower. In both cases, the creation rate for parasitic  Rb$_{2}^{+}  $, estimated based on findings reported in~\cite{Haerter2013b}, is approximately $ 4 \, \text{s}^{-1}$, taking place on much longer timescales.} In addition, Raman sideband cooling provides a method for probing the temperature of the ion with high resolution in a range where implementations based on the radial-cutoff model are insensitive. The additional control over the axial position of the ion achieved by adjusting the electrostatic potential with the endcap electrodes would then allow to move the ion into a region of low atomic density with the additional advantage of controlling the collision energy at an unprecedented level of precision~\cite{Leibfried2017}. 

As a second example for a scheme optimized for reaching collision energies close to the s-wave limit we consider a trap configuration similar to that shown in Figures \ref{fig:setup4cooling} and  \ref{fig:bichro}, again designed to trap Doppler cooled barium ions with near unity probability but with the potential tuned to give a depth of about $U_{0}^{Rb}/k_B \approx 10 \, \mu \text{K} $  for the rubidium ensemble. {This could be achieved using the experimental parameters summarized in \mbox{Table~\ref{SimpleBiODTtable}}. }%
\begin{table}[h]
	\begin{tabular}{lcc}
		\toprule
		\textbf{}	& \textbf{$ ^{138} $Ba$ ^{+} $}	& \textbf{ $ ^{87} $Rb}\\
		\hline
		$P_{VIS / NIR} \, \text{(mW)}  $		& \multicolumn{2}{c}{ 176 / 469 } \\
		$w_0 \, ( \mu \text{m})  $		& \multicolumn{2}{c}{ 3.7 }  \\
		$\omega_{ax}/(2 \pi) \, \text{(kHz)}  $		& 2.7			& 44.5\\
		$\omega_{rad}/(2 \pi) \, \text{(kHz)}  $		& 1.67			& 1.65\\
		$U_{0}/k_B \text{(K)} $ 		& $ 2.95 \times 10^{-3} $			& $ 10.5\times 10^{-6} $\\
		\toprule
	\end{tabular}
	\caption{Properties of the calculated bichromatic potentials for an adapted configuration based on the setup shown in Figure \ref{fig:setup4cooling}.
	}
	\label{SimpleBiODTtable}
\end{table}
A prerequisite for this proposed setup is a high pointing stability that could be achieved by sending in both ODT beams from the same direction and by focusing them using the same objective. For the expected trap frequencies and atom numbers around $N_{Rb} \approx 4 \times 10^{3} $ prepared at a temperature of $ T_{Rb} \approx 1 \, \mu \text{K} $, close to or below the critical temperature, we calculate a peak density of $n_0 \approx 10^{15} \, \text{cm}^{-3} $. As illustrated in \mbox{Figure \ref{Overlap}a}, under these conditions, the three-body recombination rate $ \gamma_{3b} $ would exceed the Langevin rate $ \gamma_{Lgvn} $ resulting in ion loss before reaching thermalization with the atomic gas, assuming a Ba$ ^{+} $-Rb-Rb three-body loss rate coefficient of $ k_3 = 1.03 \times 10^{-24} \, \text{cm}^{6} \text{s}^{-1}$ reported in~\cite{Kruekow2016a}. 
To achieve conditions that support sympathetic cooling it is therefore necessary to have a way to tune the ratio $ \gamma_{Lgvn} / \gamma_{3b} $. This could be realized by placing the ion at a distance $ \Delta_{IA,z} $ from the cloud center rather than shifting it into cloud regions close to the center with near peak density as done before~\cite{Leibfried2017}. Then, the effective local density probed by the ion is no longer the peak density $ n_0 $ but instead determined by the overlap of the atomic and ion density distributions, expressed as an ion--atom overlap integral $ n_{IA} = \int^{\infty}_{ - \infty} n_{atom}(z) n_{ion}(z) \, dz$. As shown in Figure \ref{Overlap}b, the axial overlap calculated for an ion prepared at $ T_{D} $ and {an atomic ensemble that for simplicity is assumed to be a thermal gas with a temperature of} 
$T_{at} = 1 \, \mu \text{K} $ (shown as black, blue and red shaded regions in the inset)  can be adjusted by more than five orders of magnitude even for comparatively small separations of less then $ 15 \, \mu \text{m}$. {For instance, with the above parameters, an ion--atom separation of $ \Delta_{IA,z}  = 11.3 \, \mu \text{m} $ would allow to thermalize with the atomic cloud within less than 10 ms, while the expected $ \gamma_{3b} \approx 67 \,  \text{s}^{-1}$ at the final temperature is much less than $ \gamma_{Lgvn} \approx 10^{3} \,  \text{s}^{-1}$.} This gives minute control over the average atomic density probed by the ion's wavefunction, and allows adjusting the ratio of Langevin collisions to three-body recombination events even for extremely low temperatures below $ 1 \, \mu \text{K} $ and high atomic densities on the order of $ 10^{15} \, \text{cm}^{-3}$ typical for quantum degenerate gases. The resulting timescales for the discussed processes are determined by several experimental parameters such as trap frequencies, thermal fraction of the atomic ensemble, initial temperatures and the investigated species. Notwithstanding, the proposed method is founded on precise control over the overlap between wavepackets, and its general effectiveness is expected to remain unaffected whenever a separation on the order of their spatial extent can be achieved.\\

Since our approach is not restricted to the combination of Ba$ ^+ $ and Rb, utilizing a different ion--atom mixture, such as $^6 $Li$ - ^{138} $Ba$ ^+$, is an option offering several advantages in view of reaching the quantum mechanically dominated regime of interaction. On the one hand, the predicted threshold temperature for observing such interactions is approximately $ 10 ~ \mu \text{K}$, more than two orders of magnitude higher than for the $ ^{87} $Rb$- ^{138}$Ba$ ^+ $ system~\cite{Cetina2012}. On the other hand, using fermions as a bath may allow to suppress three-body recombination events, enabling much longer interaction times. In addition, due to the unusually large ion-to-atom mass ratio, the impact of micromotion-induced heating is expected to be substantially mitigated~\cite{Cetina2012}, similar to the $ ^{6}\text{Li}-{}^{171}\text{Yb}^+ $ {and $ ^{6}\text{Li}-{}^{138}\text{Ba}^+ $} systems where sympathetic cooling was demonstrated recently~\cite{Feldker2020,Weckesser2021b}. This may enable pre-cooling $ \text{Ba}^+ $ to lower energies in the hybrid trap which then would allow for optical trapping at much lower laser intensities. This would have a similarly positive effect as in the proposed case of sub-Doppler cooling but without the requirement of additional lasers for driving Raman transitions or high trapping frequencies needed for operating in the resolved sideband regime~\cite{Monroe1995,Hamann1998,Vuletic1998,Karpa2013}.
\section{Conclusions}

To summarize, we have discussed the influence of beam pointing stability, as one of the currently dominant limitations to the achievable performance in the recently demonstrated approach to realizing ion--atom interactions in the quantum dominated regime based on bichromatic optical dipole traps. Our calculations show that even comparatively small fluctuations of the trap overlap on the order of a few micrometers can drastically reduce the effective atom density probed by an immersed ion. The main conclusions of our numerical analysis are twofold: firstly, the optical pointing stability is crucial for achieving thermal equilibration with the atomic ensemble, and secondly, the resolution of optical trapping probability as an observable for deriving ion temperature substantially suffers in the presence of beam fluctuations. As a consequence, the extracted temperatures from experiments averaging over different trapping beam configurations represent a lower bound on the actually achievable cooling performance. On the other hand, with an improved stability of the overlap between the optical dipole traps, collision energies are expected to approach a range where Langevin collisions compete with three-body recombination processes. To avoid rapid ion loss, we propose optimized trap configurations suitable for controlling ion--atom collisions by making use of the flexibility afforded by optical potentials and the additional degrees of freedom accessed through the electrostatic components of the effective ion trapping potential.
%

\end{document}